\documentclass{article}
\usepackage[utf8]{inputenc}

\usepackage[margin=20mm]{geometry}
\usepackage{multicol}
\usepackage[normalem]{ulem}
\usepackage{titlesec}
\titleformat{\section}
  {\bf\normalsize\filcenter}{\thesection.}
  {1ex}{}
\titleformat{\subsection}
  {\bf\normalsize}{\thesection.}
  {0ex}{}
\renewcommand{\thesection}{\Roman{section}}
\newcommand{\chapter}[1]{}

    \usepackage{graphicx}
    
    \newcommand{\breakl}{\par\noindent\rule[1ex]{0.5\textwidth}{0.4pt}}
    \newcommand{\breakr}{\begin{flushright}\par\noindent\rule[1ex]{0.5\textwidth}{0.4pt}\end{flushright}}
    \newcommand{\breakc}{\begin{center}\par\noindent\rule[1ex]{0.75\textwidth}{0.4pt}\end{center}}
    \usepackage{caption}

\usepackage{amsmath}
\usepackage{amssymb}
\usepackage{siunitx}
\newcommand{\angstrom}{\si{\angstrom}}
\newcommand{\n}{\nonumber\\}
\renewcommand{\matrix}[1]{\begin{tabular}{cccccccc}#1\end{tabular}}

\usepackage{cite}
\begin{document}
\begin{center}
{\large\bf{Chemistry beyond the Hartree-Fock limit via quantum computed moments}}
\end{center}
\begin{center}
Michael A. Jones$^{\rm1}$, 
Harish J. Vallury$^{\rm1}$, 
Charles D. Hill$^{\rm1,2}$, 
Lloyd C. L. Hollenberg$^{\rm1}$
\\\vspace{1ex}
{\it {\footnotesize $^{\rm1}$School of Physics, University of Melbourne, Parkville 3010, AUSTRALIA\\
$^{\rm2}$School of Mathematics and Statistics, University of Melbourne, Parkville 3010, AUSTRALIA}}
\end{center}
{\bf Quantum computers hold promise to circumvent the limitations of conventional computing for difficult molecular problems. However, the accumulation of quantum logic errors on real devices represents a major challenge, particularly in the pursuit of chemical accuracy requiring the inclusion of dynamical effects. In this work we implement the quantum computed moments (QCM) approach for hydrogen chain molecular systems up to H$_6$. On a superconducting quantum processor, Hamiltonian moments, $\langle \mathcal{H}^p\rangle$ are computed with respect to the Hartree-Fock state, which are then employed in Lanczos expansion theory
to determine an estimate for the ground-state energy which incorporates electronic correlations and manifestly improves on the variational result.
Post-processing purification of the raw QCM data takes the estimate through the Hartree-Fock variational limit to within 99.9\% of the exact electronic ground-state energy for the largest system studied, H$_6$. Calculated dissociation curves indicate precision at about 10mH for this system and as low as 0.1mH for molecular hydrogen, H$_2$, over a range of bond lengths. In the context of stringent precision requirements for chemical problems, these results provide strong evidence for the error suppression capability of the QCM method, particularly when coupled with post-processing error mitigation. Greater emphasis on more efficient representations of the Hamiltonian and classical preprocessing steps may enable the solution of larger systems on near-term quantum processors.
}

\begin{multicols}{2}
\section{Introduction}\label{sec:introduction}
\begin{figure*}[t!]
\begin{center}
\includegraphics[width=\linewidth]{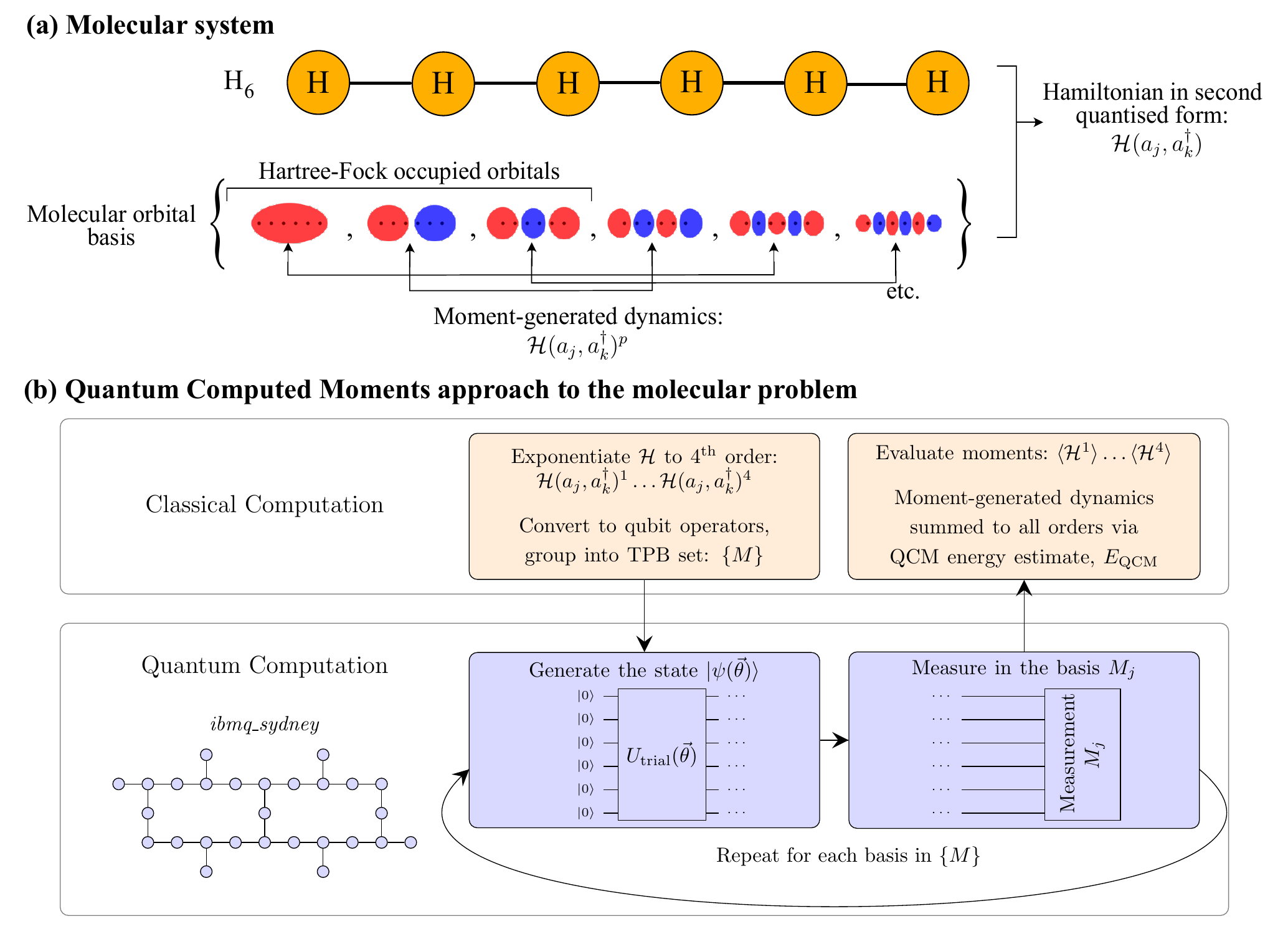}
\caption{Overview of the quantum computed moments (QCM) approach applied to problems in chemistry. 
(a) The molecular system H$_6$ is represented by a second-quantised Hamiltonian over a set of molecular orbitals. The trial-state is the Hartree-Fock state, i.e. the occupation of the indicated orbitals. Application of the Hamiltonian moments allows for the generation of dynamical effects that the Hartree-Fock state cannot otherwise incorporate.
(b) Overview of hybrid quantum/classical aspects of the QCM approach including a device map for the quantum processor \emph{ibmq\_sydney} used in this work.
}
\label{fig:overview}
\end{center}
\end{figure*}

The computing resources required for the ab-initio solution of molecular systems generally scale exponentially as the system size increases, however, Feynman recognised that quantum computers may be able to solve these problems efficiently \cite{feynman}. There has been considerable progress in developing quantum algorithmic approaches to the problem, and in understanding the quantum resources required for useful real-world cases \cite{reiher_FeMoco,montgomery_FeMoco,elfving_QChemSummary,vonBurg_catalysis,wecker_QFCIGateCounts,kivlichan_FTPeriodic}. Generally, because the underlying quantum algorithms are inherently phase sensitive, these approaches require fault-tolerant quantum error correction over hundreds of thousands to several million physical qubits to simulate molecular or condensed matter systems of scientific interest. With fault-tolerant quantum computation inaccessible in the short-to-medium term, approaches have been suggested that aim to make use of the advantages of quantum computation while keeping circuit depth minimal to reduce the accumulation of errors. Notable among these methods are variational hybrid algorithms such as the Variational Quantum Eigensolver (VQE) \cite{peruzzo_VQE,kandala_IBMVQE} which exploits the quantum processor's ability to efficiently encode the state of a quantum system while leveraging classical computation to optimise the state with respect to some inbuilt parameter set, using the expectation value of a chosen observable, usually the Hamiltonian, $\mathcal{H}$, as the cost function. Such algorithms have been considered as candidates \cite{cerezo_VQA,elfving_QChemSummary} for demonstrating quantum advantage \cite{preskill_Supremacy} on a problem of real scientific interest. Since the initial proposal \cite{peruzzo_VQE} and implementations of VQE \cite{peruzzo_VQE,shen_vqe,kandala_IBMVQE,o-malley_googleVQE}, various modifications and improvements to the algorithm, such as adaptive ans\"atze \cite{grimsley_AdaptVQE,tang_qubitAdapt} and alternative objective functions \cite{santagati_waves} have been suggested. 

While the reduced circuit depth of variational quantum algorithms provides a way to reduce errors, there is no way to completely prevent them on the Noisy Intermediate Scale Quantum (NISQ) hardware currently available \cite{preskill_NISQ}. As such, methods have been proposed to mitigate the effects of noise such as Richardson extrapolation \cite{richardson_DeferredApproach,kandala_ErrorMitigation} and McWeeny purification \cite{mcweeny_DensityMatrices,google_HF} among others. For molecular problems, these methods have achieved some success in effectively recovering the noise-free limit of the variational trial-state -- usually constructed in the Hartree-Fock approximation \cite{thouless_HFTheory} as a first step. However, the real challenge remains to incorporate dynamics (electron correlations) in a sufficiently noise-robust manner to break through the Hartree-Fock limit and into the regime of chemical precision at the 1 kcal/mol (1.59 mH) level. While a more complicated trial state such as the UCC ans\"atz \cite{taube_UCC} or its variations \cite{lee_kUpCCGSD} can, in principle, incorporate the required correlations; as the problem size is increased beyond what can be simulated classically the circuits soon become prohibitively long and inaccessible to near-term quantum devices.

Here we apply the recently introduced Quantum Computed Moments (QCM) approach \cite{vallury_QCM} to compute dynamical corrections to the variational estimate for molecular problems on a superconducting quantum processor as outlined in Figure \ref{fig:overview} and discussed in further detail in Section \ref{sec:method}. The QCM approach incorporates system dynamics through the computation of Hamiltonian moments, and uses results from Lanczos expansion theory \cite{hollenberg_PlaquetteExpansion} to produce a dynamic correction to the variational result, effectively summing these effects to all orders. The utility of the QCM method was previously demonstrated for quantum magnetism problems on a superconducting quantum processor of up to 25 qubits, providing stable estimates of the ground-state energy which improve on the corresponding variational results. Critically, the QCM results, even for this relatively large number of qubits, showed a high level of robustness to device errors and noise, suggesting utility for other quantum problems of interest on near-term devices. For our test molecular problem, we consider the ground state energy of hydrogen-atom chains up to H$_6$ computed with respect to the Hartree-Fock variational state.

Recently, other methods of using moments in the QC context have been proposed based on the power method \cite{seki_QuantumPowerMethod}, extensions of the variational approach \cite{suchsland_ThirdMoment}, or the connected moments expansion (CMX) \cite{kowalski_CMX,peng_VariationalPDS,claudino_cmx,cioslowski_CMX}. Unlike approaches such as the CMX, which applies Pade approximates to the $t$-expansion, the energy estimates obtained from Lanczos expansion theory are based on a rigorous diagonalisation of the Hamiltonian in Lanczos expanded form for a given finite moment order. In the past, direct comparisons show that the Lanczos expansion approach consistently provides a better energy estimate \cite{hollenberg_AnalyticSolution}.

These moment-based methods fit into a broader category of so-called Quantum Subspace Expansion (QSE) methods that classically diagonalise the Hamiltonian in an alternative basis where the diagonalisation can be carried out more efficiently/accurately. The generation of such a basis can be carried out by repeated application of the Hamiltonian operator to a suitably chosen trial-state (leading to moments-based methods) or by the application of other operators such as electronic excitations \cite{mcclean_qse,takeshita_qse}, Pauli operators \cite{colless_qse} or matrix exponentials of the Hamiltonian \cite{motta_QLanczos,yeter-aydeniz_QLanczos}. Alternatively, a non-orthogonal basis can be defined as a more general set of states that can be prepared easily on the quantum processor \cite{huggins_novqe,stair_mrsqk,parrish_QFD,cohn_QFD}.

It is worth noting that Lanczos expansion theory is capable of more than calculating corrections to the ground state energy and that it should also be possible to calculate energy gaps \cite{hollenberg_MassGap}, thermodynamic properties and expectation values of physical quantities \cite{kassal_HellmannFeynman} etc. Recently \cite{jamet_KVQA} a simulated quantum computer was used to calculate Green's functions based on the Hamiltonian moments and \cite{GuzmanLacroix_moments} details a method for calculating exited states and time-evolution over a long time period from the Hamiltonian moments obtained from time-evolution over a short time period.

\section{Results}

\subsection*{Chemistry via Quantum Computed Moments}\label{sec:method}
\begin{figure*}[tp]
\includegraphics[width=\linewidth]{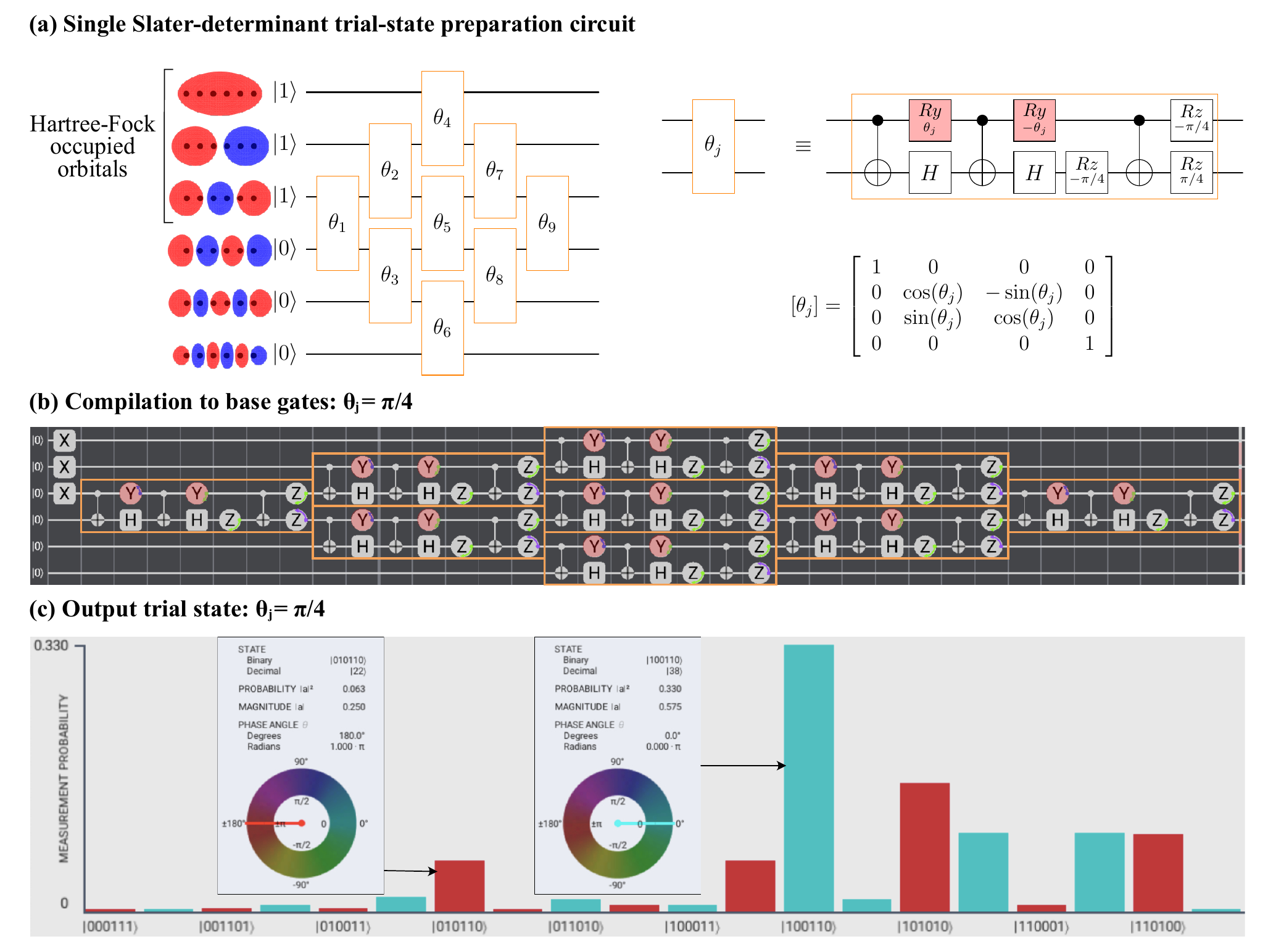}
\caption{Trial state construction and implementation: illustrated case -- the 6 atom hydrogen chain. 
a) Qubit representation of the orbital basis, the trial circuit expressed in terms of Givens rotations \cite{wecker_SolvingElectronModels,kivlichan_LinearDepth,arrazola_GivensCircuits} and the Givens rotation expressed in terms of CNOT and single qubit gates and its matrix representation. Parameterised gates are shaded in red 
b) The trial circuit for H$_6$ with each parameter set to $\theta_j=\pi/4$, compiled to the gate set \{H, X, Ry, Rz, CNOT\}. Parameterised gates are shaded in red and each block implementing a Givens rotation is outlined in orange. 
c) The trial state produced by the circuit in c). Only the states representing the correct number of electrons are included on the horizontal axis since only these states can have non-zero amplitudes. 
b) and c) are visualised using the Quantum User Interface (QUI) system \cite{qui}
}
\label{fig:circuit}
\end{figure*}

The QCM method is applicable to any molecular problem in general, however, for definiteness we consider linear chains of hydrogen atoms (Figure 1a) governed by the usual second-quantised molecular Hamiltonian:
\begin{align}
\mathcal{H}&=\sum_{jk}t_{jk}a^\dagger_ja_k+\sum_{jklm}t_{jklm}a^\dagger_ja^\dagger_ka_la_m,\label{eqn:hamiltonian}
\end{align}
where the one- and two-body molecular integrals, $t_{jk}$ and $t_{jklm}$ respectively, are computed efficiently on a classical computer
. The Hamiltonian as written in Equation \ref{eqn:hamiltonian} does not include the repulsion energy between the atomic nuclei -- when calculating dissociation curves this must be taken into account as the nuclear repulsion varies as a function of atomic coordinates. This energy is referred to as the molecular energy. When performing optimisations for a single set of atomic positions the nuclear repulsion energy does not need to be taken into account as it is independent of the electronic configuration -- this energy is referred to as the electronic energy. The problem is to use a quantum computer to find the ground-state energy to chemical precision (1 kcal/mol $\approx$ 1.59 mH). The conventional Variational Quantum Eigensolver (VQE) approach employs the quantum computer to compute the expectation value $\langle\mathcal{H}\rangle$, with respect to a well-chosen trial state, as a cost function in a classical minimisation loop. For chemical problems, the Hartree-Fock (HF) state is the chosen starting point for the minimisation of $\langle\mathcal{H}\rangle$. While gate-errors and device noise have a considerable effect on the values of $\langle\mathcal{H}\rangle$, error mitigation and purification techniques can essentially recover the HF variational energy \cite{google_HF}. In the quantum computation context, to go beyond the HF limit towards chemical precision one must include electronic correlations. While this can be achieved through the use of better trial state ans\"atz such as the Unitary Coupled Cluster ans\"atz (UCC) \cite{taube_UCC} or alternative algorithms such as Quantum Phase Estimation (QPE) \cite{kitaev_qpea}, these approaches generally involve many qubits and/or deep circuits and/or quantum error corrected logical qubits and are therefore not suited to NISQ devices, even in the medium term. For NISQ devices to be of any use as a computational tool for chemistry, the algorithmic approaches must be adapted to be highly noise-robust in order to be capable of producing accurate results in the context of chemical precision.

The QCM method is based on an expansion of the Lanczos tridiagonal form in terms of moments $\langle\mathcal{H}^p\rangle$ (see section \ref{sec:qcm} in SI for details) \cite{hollenberg_PlaquetteExpansion}. The Hamiltonian moments encapsulate the system's dynamics with respect to a given trial state -- for the chemistry problems considered here, this equates to incorporating electronic correlations over the single Slater determinant trial state. In the quantum computing context, early explorations were conducted in \cite{duan_MPS} and later extended to the notion of direct computation in \cite{jones_qcm}. The resulting Quantum Computed Moments (QCM) method \cite{vallury_QCM} employs an approximation for the ground state energy in terms of connected moments (cumulants) $c_p$ of $\langle \mathcal{H}^p\rangle$, to fourth order, given by the expression \cite{hollenberg_LatticeHamiltonians}:
\begin{equation}
E_{\rm QCM}\equiv E_0^{(4)}=c_1-\frac{c_2^2}{c_3^2-c_2c_4}\left(\sqrt{3c_3^2-2c_2c_4}-c_3\right).\label{eqn:Eqcm}
\end{equation}
The second term involving higher order connected moments not only provides a dynamical correction to the variational result, $c_1\equiv\langle\mathcal{H}\rangle$, it also contributes a high degree of robustness to circuit errors.

For a given Hamiltonian, $\mathcal{H}$, the QCM method begins by exponentiating $\mathcal{H}$ to produce $\{\mathcal{H}^1, \mathcal{H}^2, \mathcal{H}^3, \mathcal{H}^4\}$. After conversion from second-quantised form to qubit operators, the growth of the number of terms in the exponentiated forms of $\mathcal{H}$ is controlled by forming tensor product basis (TPB) sets \cite{kandala_IBMVQE} in a classical pre-processing step (as discussed in section \ref{sec:scaling} of the Supplementary Information). The trial state employed in this work is based on that used in reference \cite{google_HF}, where each qubit represents the occupation state of a molecular orbital, classically pre-computed in the STO-3G minimal basis using the \emph{Python} package \emph{pyscf} \cite{pyscf}. (See section \ref{sec:circuit} in the Supplementary Information for additional details.)

To reduce the computational burden on the quantum processor an extension of the spin-symmetry reduction of \cite{google_HF} was employed to reduce the number of qubits by a factor of 2. While this qubit reduction (see section \ref{sec:reduction} in SI for details) requires several restrictions on the problem, most notably that the trial state must be a single Slater-determinant, the inclusion of electron dynamics introduced to the system by the QCM method allows the computation to achieve accuracy beyond what would normally be possible for such a trial state. This technique also allows for a reduction in the number of Tensor Product Basis (TPB) elements that need to be measured in the Hamiltonian averaging procedure by a factor of $N_s^{14}$ from the naive method of measuring the (worst case) $\mathcal{O}(N_s^{16})$ terms in $\mathcal{H}^4$ individually, where $N_s$ is the number of spin-orbitals. The results here were obtained using the conventional $\mathcal{O}(N_s^4)$ scaling of the Hamiltonian terms for which classical pre-processing on modest computing resources limited the molecular system size. However, our results together with the wide applicability of the QCM method to molecular systems in general, provides the tantalising possibility that with alternative Hamiltonian representations the QCM approach, coupled with error mitigation schemes, has the potential to provide accurate and strongly error-robust results for the ground state energy of larger chemical systems on near-term quantum computers.

\subsection*{Circuit parameter sweep}\label{sec:results}
\begin{figure*}[t]
\includegraphics[width=\linewidth]{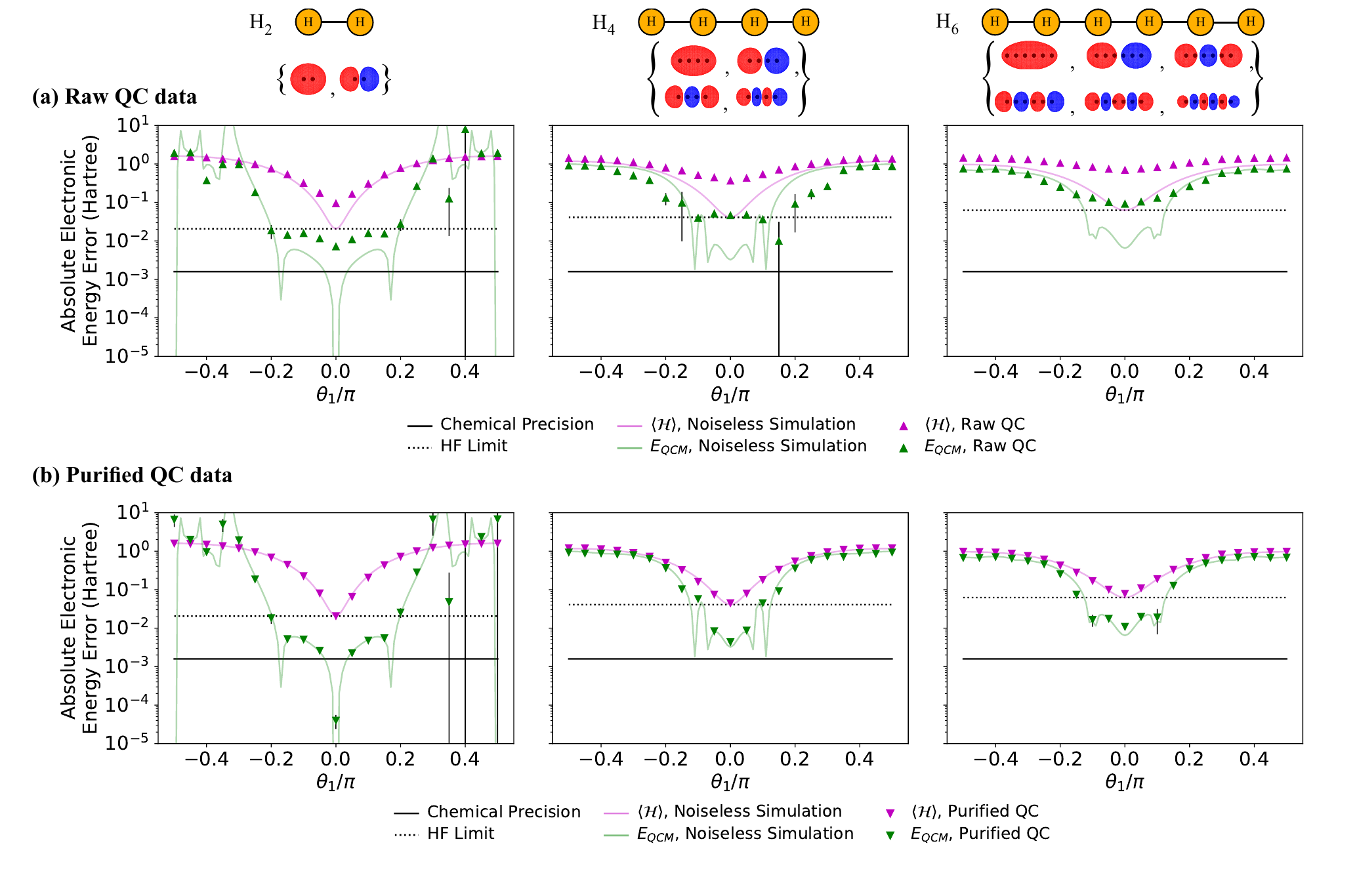}
\caption{Results of varying the leading circuit parameter, $\theta_1$ for (from left to right) H$_2$, H$_4$, H$_6$. The vertical axis is the absolute error in electronic energy from the FCI result $|E-E_{\rm FCI}|$ on a logarithmic scale. Ideal simulations were performed without errors or shot noise while QC data was averaged over 4 runs (total 8192 shots). Statistical error bars on the QC data represent one standard deviation and are often smaller than the data points.
a) Comparison of raw data results to noiseless simulation. b) Comparison of purified data to noiseless simulation.
}
\label{fig:results}
\end{figure*}
\begin{figure*}[t]
\includegraphics[width=\linewidth]{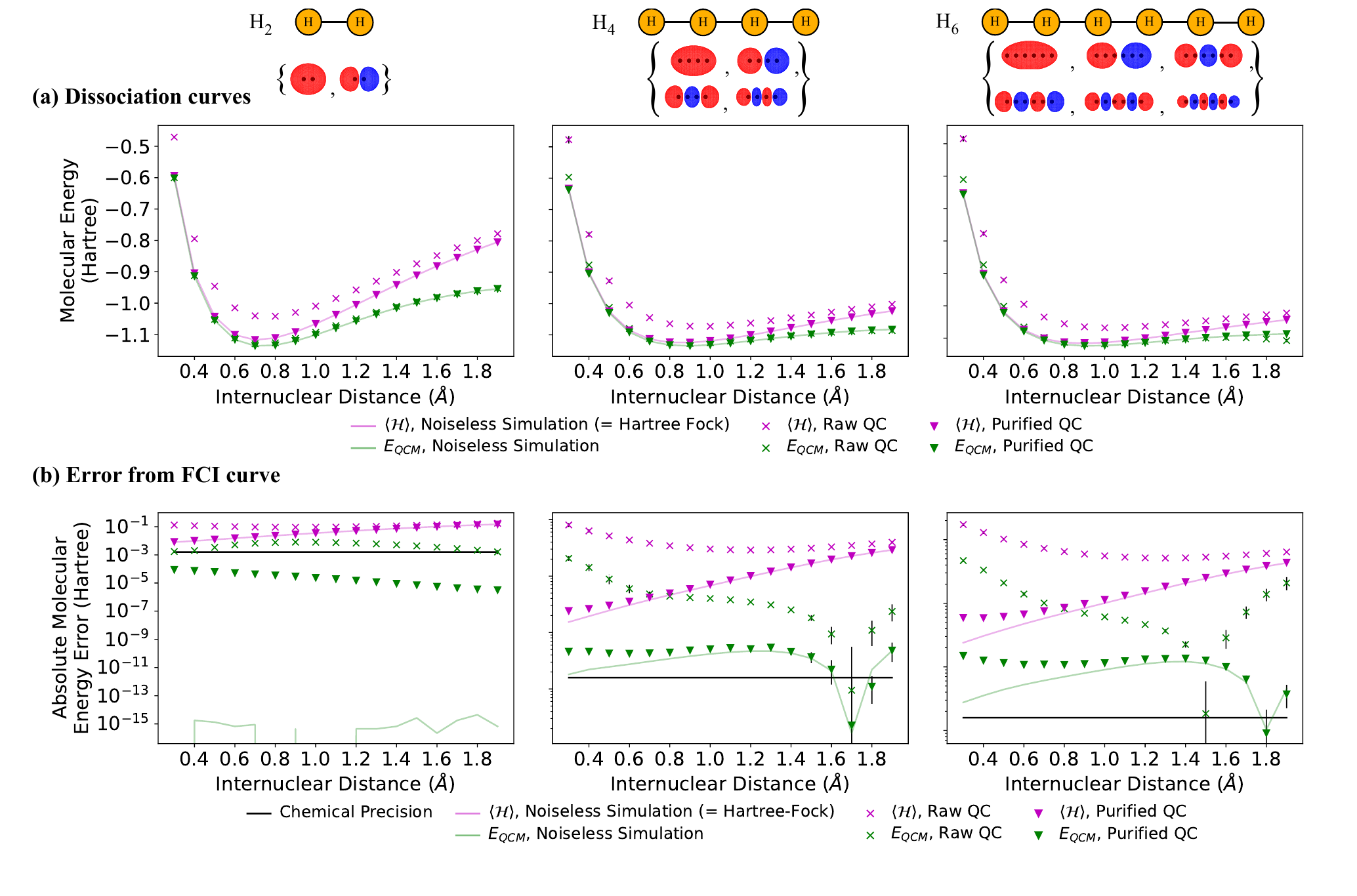}
\caption{Dissociation curves for hydrogen chains. The energy values at each point correspond to a single energy evaluation (8192 shots) performed at parameter value $\theta=(0,0,\dots)$. a) Molecular energy as a function of inter-nuclear distance. The purple line is the noiseless simulation result which reproduces exactly the Hartree-Fock limit. b) The absolute value of the error in molecular energy relative to the FCI results.}
\label{fig:dissociation}
\end{figure*}

With respect to the single Slater-determinant trial-state defined in Figure \ref{fig:circuit} (parameter set $\vec{\theta}$), all observables contained in the 
TPB sets were measured directly on the {\it ibmq\_sydney} device for the molecular systems H$_2$, H$_4$ and H$_6$ with all bond lengths set to 0.74$\angstrom$ (the equilibrium bond distance of molecular hydrogen). Each TPB set was measured using 8192 shots. To illustrate variational behaviour we present the data for values of the opening trial-state parameter $\theta_1$ around the Hartree-Fock state compared to the Hartree-Fock limit and the Full Configuration Interaction (FCI) results in Figure \ref{fig:results}a) (see section \ref{sec:SIdata} in the Supplementary Information for ensemble data over the variational parameter sets).
The error robustness of the QCM results is quite evident for all three molecular systems -- while the variational result (with respect to $\theta_1$) moves significantly away from the trial-state (HF) limit, the upward shift of the QCM due to device errors is suppressed by an order of magnitude. Despite the large number of observables required to be measured on the QC device in the determination of the moments, the QCM correction term consistently suppresses the device errors contained in the variational computation $c_1 = \langle \mathcal{H}\rangle$. In fact, for the largest system, H$_6$, the raw QCM data already recovers 99.7\% of the minimised trial-state energy at the HF point.

\subsection*{Density matrix purification}
To investigate the potential of the method to reach beyond the Hartree-Fock limit we perform McWeeny purification on the 1-body reduced density matrix (1-RDM) \cite{mcweeny_DensityMatrices} -- this procedure has been shown to improve the variational result at the Hartree-Fock point \cite{google_HF}. Details on the purification procedure can be found in section \ref{sec:purification} in the Supplementary Information. The results of the RDM purification procedure are shown in Figure \ref{fig:results}b). As observed in \cite{google_HF}, the purification procedure applied to the variational result goes a long way to recovering the Hartree-Fock limit. However, the key observation is that the RDM purification is also highly effective in correcting the remaining device errors in the QCM calculation, explicitly recovering the system dynamics and pushing the result beyond the HF limit. 

\subsection*{Dissociation curves}
Finally, we repeat the calculations at a range of atomic separations to compute the dissociation curves shown in Figure \ref{fig:dissociation}. As expected, in the absence of noise, the QCM correction incorporates the electronic correlations at a sufficient level to recover energies below the Hartree-Fock limit for all bond distances examined for all three molecules. 
For most cases, especially at longer bond lengths where the Hartree-Fock limit diverges more significantly from the FCI results, the moments-based correction without any additional error mitigation is sufficient to calculate energies below the Hartree-Fock limit, even when performed on a noisy quantum device.

With the application of 1-RDM purification, the moments based method on the quantum device \emph{ibmq\_sydney} was able to outperform the noiseless Hartree-Fock results for all molecular geometries considered, reaching error thresholds relative to the FCI results of order 10 mH for the longest chain, H$_6$, and 0.1 mH for molecular hydrogen. The latter is below the chemical precision threshold of $1.59$ mH. We note that this result is within chemical precision of the FCI energy calculated in the minimal STO-3G basis and that a significantly larger basis is required to claim true chemical accuracy \cite{elfving_QChemSummary}. We also note that the results presented in \cite{google_HF} are likewise restricted by the choice of basis and are in fact within chemical precision only of the Hartree-Fock limit and not the FCI ground state
. Additionally, the results presented here do not rely on the symmetry of the system and could, in principle, be calculated for any spatial arrangement of the atoms. This is in contrast to the results of \cite{kawashima_H10} where chemical precision relative to FCI calculations was achieved for the H$_{10}$ ring by exploiting the high level of rotational symmetry in the system. Other results reporting chemical precision include work reported in \cite{mccaskey_AlkaliHydrides} and \cite{nam_WaterIonTrap}: the former simulates NaH on 4 superconducting qubits using a frozen core approximation and a UCC inspired circuit, a problem of similar size to the H$_2$ molecule simulated here, while the latter simulates H$_2$O on 11 trapped-ion qubits by carefully selecting the excitation operators to be included in a UCC-style ans\"atz circuit. Given the precision of the QCM method when performed using the less accurate single Slater-determinant trial state it is expected that application of the method to UCC-style ans\"atz circuits would provide a further improvement on the precision of the ground-state energy estimates.

\section{Discussion}\label{sec:conclusion}
We have applied the Quantum Computed Moments method to the quantum-chemical problem of computing the ground state energy of linear chains of hydrogen atoms and found that, even for restrictive single Slater-determinant trial states, use of the QCM method allows for recovery of electron correlation and therefore of energies below the Hartree-Fock threshold. Though computation of the Hamiltonian moments may seem expensive, the scaling of the number of terms in $\mathcal{H}^4$ is significantly better than the worst-case as seen in appendix \ref{sec:scaling} and could be further reduced by transformation into an alternate basis. Additionally the number of measurements can be controlled by grouping mutually commuting operators for simultaneous measurement.

With the addition of McWeeny purification, we demonstrate that the method is capable of outperforming (noise-free) Hartree-Fock calculations, even when the moments are computed on noisy present-day quantum hardware, for chains of up to 6 atoms, the largest system studied here requiring up to 27 CNOT gates. For molecular hydrogen, H$_2$, we achieve results that are within chemical precision of the FCI result calculated with respect to the minimal basis set for a range of inter-nuclear distances around the equilibrium bond length. For the H$_6$ chain at 0.74\angstrom, with no error mitigation the QCM method is able to recover 97.1\% of the molecular energy while the usual direct measurement of $\langle\mathcal{H}\rangle$ is able to return only 78\% of the energy due to noise in the trial circuit preparation.

Although the McWeeny purification technique is not easily generalised to states that cannot be represented as a single Slater-determinant, the QCM method performs well even without the purification and would likely benefit from any improvement in the trial state, due to its significant robustness to noise. Furthermore it is possible to adapt the QCM method for the computation of excited state energies \cite{hollenberg_MassGap} and to properties other than the energy \cite{kassal_HellmannFeynman}. With improvement of the trial state, combined with error mitigation techniques and alternative Hamiltonian representations to control the scaling of the problem the QCM method is a promising technique for the pursuit of chemical accuracy on present-day quantum hardware.

\section{Methods}

\subsection*{Spin-degeneracy qubit-reduction}
Due to the spin-symmetry of the system the number of qubits required for quantum simulation can be reduced by a factor of 2 if the trial state is restricted to a single Slater-determinant, $|\Psi_P\rangle$ (see section \ref{sec:reduction} in SI). Given the restricted trial state it is possible to extract expectation values for any excitation operator from the 1-body reduced density matrix (1-RDM) according to the equation
\begin{align}
\langle \Psi_P|a^\dagger_ja^\dagger_k\dots a_{k'}a_{j'}|\Psi_P\rangle 
&=\left|\matrix{
$\langle a^\dagger_ja_{j'}\rangle$&$\langle a^\dagger_ja_{k'}\rangle$&$\cdots$\\
$\langle a^\dagger_ka_{j'}\rangle$&$\langle a^\dagger_ka_{k'}\rangle$&$\cdots$\\
$\vdots$&$\vdots$&$\ddots$
}\right|.
\end{align}
The qubit-reduction method is described in more detail in section \ref{sec:reduction} in the Supplementary Information.

\subsection*{The chemical Hamiltonian}
The chemical Hamiltonians used in this work were computed using the \emph{python} package \emph{pyscf} \cite{pyscf} to optimise the molecular orbitals and were converted to qubit Hamiltonians using the Jordan-Wigner transform \cite{JordanWigner_JWTransform}.

\subsection*{Measurement details}
To reduce the required number of state preparations the Pauli strings required for measurement of the 1-RDM are grouped into $\mathcal{O}(N_s^2)$ mutually commuting tensor product basis sets \cite{kandala_IBMVQE} as described in section \ref{sec:scaling} in the Supplementary Information. For each TPB, the trial circuit was executed and the output state was measured 8192 times to obtain the average results. To estimate uncertainties, the 8192 measurement results were randomly assigned to one of 4 bins. Each bin (of roughly 2000 results each) was processed individually and the standard deviation of these 4 results was calculated. Quantum computed data for the graphs in figures \ref{fig:results} and \ref{fig:dissociation} were taken from the device \emph{ibmq\_sydney} and simulated results were calculated using a statevector simulator without shot noise. Dissociation curves were also calculated on the \emph{ibmq\_toronto} and \emph{ibmq\_guadalupe} devices and found to be consistent with the \emph{ibmq\_sydney} results.

\subsection*{Density matrix purification}
Purification of the 1-body reduced density matrix was performed following \cite{mcweeny_DensityMatrices,google_HF} by iteratively applying the equation:
\begin{align}
R_{j+1}=3R_j^2-2R_j^3,
\end{align}
where $R_0$ is the unpurified matrix and $R_j$ is the matrix after $j$ iterations of the purification procedure. The purification method is described in more detail in section \ref{sec:purification} of the Supplementary Information.

\section{Contributions and Acknowledgements}\label{sec:ack}
LCLH conceived the project. MAJ set up the computational framework and performed the calculations, with inputs from all authors. The research was supported by the University of Melbourne through the establishment of the IBM Q Network Hub at the University. MAJ is supported by the Australian Commonwealth Government through a Research Training Program Scholarship. Classical pre- and post-processing for the H$_6$ molecule were performed using the Spartan HPC platform \cite{spartan}.
\end{multicols}

\breakc
\begin{multicols}{2}
\renewcommand{\section}[2]{}
{\scriptsize\bibliography{QChem_paper.bib}}
\end{multicols}

\newpage
\appendix
\renewcommand{\thesection}{}
\renewcommand{\section}[1]{
	\par\vspace{2ex}\par\noindent
	\begin{center}
	{\bf\large\thesection~#1}
	\end{center}}
\section{Supplementary Information}
\renewcommand{\theequation}{SI.\arabic{equation}}
\setcounter{equation}{0}
\renewcommand{\thefigure}{SI.\arabic{figure}}
\setcounter{figure}{0}
\renewcommand{\thetable}{SI.\arabic{table}}
\setcounter{table}{0}
\renewcommand{\thesubsection}{\Alph{subsection}}
\titleformat{\subsection}
  {\bf\normalsize\filcenter}{\thesubsection.}
  {1ex}{}

\begin{multicols}{2}

\subsection{The Quantum Computed Moments approach}\label{sec:qcm}
We outline the background to the QCM method as follows. For a given Hamiltonian, the Lanczos recursion tri-diagonalises the system starting with an appropriate initial trial-state, $|v_1\rangle$,
\begin{align}
|v_{i+1}\rangle &=\frac{1}{\beta_i}\left[(\mathcal{H}-\alpha_i)|v_i\rangle -\beta_{i-1}|v_{i-1}\rangle \right],
\end{align}
where $\alpha_i=\langle v_i|\mathcal{H}|v_i\rangle$ and $\beta_i=\langle v_{i+1}|\mathcal{H}|v_i\rangle$ are the diagonal and off-diagonal elements of the tri-diagonal matrix respectively \cite{lanczos_Lanczos}. By introducing the cumulants, $c_p$ \cite{horn_tExpansion,hollenberg_PlaquetteExpansion}, of the Hamiltonian with respect to the trial state $|v_1\rangle$,
\begin{align}
c_p&=\langle \mathcal{H}^p\rangle -\sum_{j=0}^{p-2}\left(\matrix{$p-1$\\$j$}\right)c_{j+1}\langle \mathcal{H}^{p-1-j}\rangle ,
\end{align}
a general expansion for the matrix elements was uncovered \cite{hollenberg_PlaquetteExpansion},
\begin{align}
\alpha(z)&=c_1+z\frac{c_3}{c_2}+z^2\frac{3c_3^2-4c_2c_3c_4+c_2^2c_5}{4c_2^4}+\dots,\n
\beta^2(z)&=zc_2+z^2\frac{c_2c_4-c_3^2}{2c_2^2}+\dots.
\end{align}
Here the continuous, positive parameter $z$ is related to the recursion index \cite{hollenberg_PlaquetteExpansion,hollenberg_AnalyticSolution,hollenberg_ClusterExpansion}. These general expressions for the matrix elements allow 
for the expression of the ground state energy in terms of an infimum theorem \cite{hollenberg_AnalyticSolution};
\begin{align}
E_0&=\inf_{z>0}\left[\alpha(z)-2\beta(z)\right].
\end{align}
Truncating the $z$-expansion 
in terms of the moments to $p_{\rm max}=4$ produces the expression for the ground state energy approximate used in the QCM method \cite{hollenberg_LatticeHamiltonians},
\begin{align}
E_{\rm QCM}\equiv E_0^{(4)}=c_1-\frac{c_2^2}{c_3^2-c_2c_4}\left(\sqrt{3c_3^2-2c_2c_4}-c_3\right),
\end{align}
where the first term is simply $c_1=\langle \mathcal{H}\rangle$ and the second term, depending on the higher order moments, provides a correction to this energy.

\subsection{Spin-degeneracy qubit reduction}\label{sec:reduction}
Although trial state depth is potentially the most precious quantum resource in a VQE calculation, any reduction in the number of qubits required for the simulation, such as through qubit tapering \cite{bravyi_QubitTapering}, will allow reduction of the circuit depth since fewer gates will be needed to generate a sufficiently entangled quantum state over the reduced number of qubits. In addition, any reduction in the number of qubits will likely decrease the number of measurements and therefore, the number of trial state preparations required per energy estimation.

There are several properties of the systems studied here, the linear hydrogen atom chains, that allow the reduction technique used in \cite{google_HF} to be applied. It should be noted that these restrictions apply only to the qubit-reduction technique not to the QCM method in general. The restrictions are;
\begin{itemize}
\item The Hamiltonian of the system is not explicitly dependant on spin, while this is true for all molecules in a vacuum, it is not true in the presence of an external magnetic field, which could potentially limit the applicability of this technique.
\item The trial state used to simulate the system consists of a single Slater determinant. The Slater determinant is often used in quantum chemistry as a starting point from which more accurate states can be constructed. However, as the objective of the QCM method is to reduce the complexity of the trial state, the use of a single determinant trial state is a reasonable choice. Additionally, it is seen that the use of the QCM method allows for the recovery of energies below the Hartree-Fock limit.
\item The system is simulated in a singlet state, while this can be done for any molecular system, the ground state of paramagnetic materials, such as oxygen gas ($O_2$), is not a singlet. In these cases, application of the qubit reduction would lead to reduced accuracy.
\end{itemize}

The Slater determinant trial state can be written in second-quantised form as;
\begin{align}
|\Psi_{\rm trial}\rangle &=|\psi_P\rangle =\prod_{p\in P}b^\dagger_p|0\rangle ,
\end{align}
where $P$ is a list of the basis states from which the determinant is constructed, $b^\dagger_p$ is the fermionic creation operator for state $p$ and $|0\rangle$ is the vacuum state. In this form, the antisymmetrisation of the state is encoded in the anticommutation of the fermionic operators. The single excitation operator $b^\dagger_qb_r$ will have vanishing expectation value with respect to the trial state unless $qr\in P$, in which case
\begin{align}
\langle\Psi_P|b^\dagger_qb_r|\Psi_P\rangle
&=\delta_{qr}\langle \Psi_P|\Psi_P\rangle -\langle \Psi_P|b_rb^\dagger_q|\Psi_P\rangle \n
&~~~~~~~=\delta_{qr},\label{eqn:2-mode-b}
\end{align}
where the last step uses $q\in P$ and $(b^\dagger_q)^2=0$ to show that the second term vanishes. From this point on, it will be assumed that 
expectation values are taken with respect to the trial state $|\Psi_P\rangle$. Using the anticommutation relations it can similarly be shown that the expectation value of a double-excitation operator is;
\begin{align}
\langle b^\dagger_qb^\dagger_rb_{r'}b_{q'}\rangle 
&=\delta_{qq'}\delta_{rr'}-\delta_{rq'}\delta_{qr'}
=\left|\matrix{$\delta_{qq'}$&$\delta_{qr'}$\\
$\delta_{rq'}$&$\delta_{rr'}$}\right|.
\end{align}
and an arbitrary level excitation operator has the expectation value;
\begin{align}
\langle b^\dagger_qb^\dagger_rb^\dagger_s\dots b_{s'}b_{r'}b_{q'}\rangle &=
\left|\matrix{
$\delta_{qq'}$&$\delta_{qr'}$&$\delta_{qs'}$&$\cdots$\\
$\delta_{rq'}$&$\delta_{rr'}$&$\delta_{rs'}$&$\cdots$\\
$\delta_{sq'}$&$\delta_{sr'}$&$\delta_{ss'}$&$\cdots$\\
$\vdots$&$\vdots$&$\vdots$&$\ddots$
}\right|.\label{eqn:detofdeltas}
\end{align}
It is worth noting that;
\begin{itemize}
\item Swapping two creation (annihilation) operators is equivalent to swapping two rows (columns) in the determinant. Both actions introduce a factor of $-1$ as expected.
\item If two creation (annihilation) operators act on the same state, then two rows (columns) in the determinant are equal and both sides of the equation vanish.
\item If a creation (annihilation) operator does not act on a state in the Slater determinant, the equation does \emph{not} hold. In this case, a modified form of the equation is required;
\begin{align}
\langle b^\dagger_qb^\dagger_r\dots b_{r'}b_{q'}\rangle &=
\left|\matrix{
$\langle b^\dagger_qb_{q'}\rangle$&$\langle b^\dagger_qb_{r'}\rangle$&$\cdots$\\
$\langle b^\dagger_rb_{q'}\rangle$&$\langle b^\dagger_rb_{r'}\rangle$&$\cdots$\\
$\vdots$&$\vdots$&$\ddots$
}\right|.
\end{align}
When all indices are $\in P$, equation \ref{eqn:2-mode-b} reduces this to equation \ref{eqn:detofdeltas}, but if any creation (annihilation) index is $\notin P$, then every entry in that row (column) will be 0 and the determinant will vanish, as required.
\end{itemize}

To allow for a more general state preparation, the basis states used in the construction of the Slater determinant are allowed to vary from the basis in which the second-quantised Hamiltonian is written. The trial state basis will be represented by the creation (annihilation) operators $b^\dagger$ ($b$) while the Hamiltonian basis will be represented by the operators $a^\dagger$ ($a$). The relation between these bases can be written as;
\begin{align}
a^\dagger_j&=\sum_qA_{jq}b^\dagger_q,
\end{align}
for appropriately normalised complex amplitudes $A_{jq}$. The expectation value for an excitation operator in the Hamiltonian basis is then
\end{multicols}

\breakl
\begin{align}
\langle \Psi_P|a^\dagger_ja^\dagger_k\dots a_{k'}a_{j'}|\Psi_P\rangle 
&=\sum_{qr\dots r'q'}A_{jq}A_{kr}\dots A^*_{k'r'}A^*_{j'q'}\langle \Psi_P|b^\dagger_qb^\dagger_r\dots b_{r'}b_{q'}|\Psi_P\rangle ,\n
&=\sum_{qr\dots r'q'}A_{jq}A_{kr}\dots A^*_{k'r'}A^*_{j'q'}\left|\matrix{
$\langle b^\dagger_qb_{q'}\rangle$&$\langle b^\dagger_qb_{r'}\rangle$&$\cdots$\\
$\langle b^\dagger_rb_{q'}\rangle$&$\langle b^\dagger_rb_{r'}\rangle$&$\cdots$\\
$\vdots$&$\vdots$&$\ddots$
}\right|,\n
&=\left|\matrix{
$\sum_{qq'}A_{jq}A^*_{j'q'}\langle b^\dagger_qb_{q'}\rangle$&$\sum_{qr'}A_{jq}A^*_{k'r'}\langle b^\dagger_qb_{r'}\rangle$&$\cdots$\\
$\sum_{rq'}A_{kr}A^*_{j'q'}\langle b^\dagger_rb_{q'}\rangle$&$\sum_{rr'}A_{kr}A^*_{k'r'}\langle b^\dagger_rb_{r'}\rangle$&$\cdots$\\
$\vdots$&$\vdots$&$\ddots$
}\right|,\n
&=\left|\matrix{
$\langle a^\dagger_ja_{j'}\rangle$&$\langle a^\dagger_ja_{k'}\rangle$&$\cdots$\\
$\langle a^\dagger_ka_{j'}\rangle$&$\langle a^\dagger_ka_{k'}\rangle$&$\cdots$\\
$\vdots$&$\vdots$&$\ddots$
}\right|.\label{eqn:qubit-reduction}
\end{align}
\breakr

\begin{multicols}{2}
Since the Hamiltonian contains only single- and double-excitations, the qubit reduction as applied in \cite{google_HF} required only the expression for double-excitation operators, which had previously been determined in \cite{mcweeny_DensityMatrices}. The more general form here is required for use with the Hamiltonian moments which can include up to 8th level excitation operators.

Equation \ref{eqn:qubit-reduction} allows the measurement of an $i$th level excitation operator to be reduced to the measurement of $i^2$ single-excitation operators and the calculation of the determinant of an $i\times i$ matrix. Since the system is being simulated in the singlet state and the Hamiltonian does not depend explicitly on the spin, any spin crossing 2-mode terms must have vanishing expectation values and all down-spin terms will have expectation values identical to the corresponding spin-up terms:
\begin{align}
\langle a^\dagger_{(j,\uparrow)}a_{(k,\downarrow)}\rangle &=0,\n
\langle a^\dagger_{(j,\uparrow)}a_{(k,\uparrow)}\rangle &=\langle a^\dagger_{(j,\downarrow)}a_{(k,\downarrow)}\rangle .\label{eqn:spin-reduction}
\end{align}
Using equations \ref{eqn:qubit-reduction} and \ref{eqn:spin-reduction} any expectation value required for estimating the Hamiltonian can be calculated by simulating and measuring only the spin-up states (or equivalently only the spin-down states), allowing the number of qubits required for simulation to be reduced by a factor of 2.

While this procedure may seem expensive at first glance, it is important to note that the level of the expectation values and therefore the size of the determinants to be computed is dependant only on the level of the moments correction used and not on the system size. Additionally, there are only $N_s^2/4$ single-excitation expectation values from which the determinants can be constructed, for a simulation involving $N_s$ spin-orbitals. Each of these expectation values only needs to be measured once and section \ref{sec:scaling} discusses ways in which the number of trial state preparations can be reduced further. From these $\mathcal{O}(N_s^2)$ expectation values, the 1-body reduced density matrix can be constructed and all non-vanishing fermionic expectation values can be written as minors (determinants of a submatrix) of the 1-RDM.

\subsection{The trial-state}\label{sec:circuit}

When applying the VQE algorithm to examples from quantum chemistry it makes sense to map the molecular orbitals, constructed as linear combinations of atomic orbitals (LCAOs), to each qubit. The molecular integrals of Equation \ref{eqn:hamiltonian} can then be calculated with respect to these orbitals. In this work, the STO-3G basis is used to represent each atomic orbital by a linear combination of 3 Gaussian orbitals, the molecular orbitals and their corresponding integrals are then calculated using the python package \emph{pyscf} \cite{pyscf} through the \emph{qiskit} package \cite{qiskit}. Because the molecular orbitals are pre-optimised by \emph{pyscf}, the solution to the Hartree-Fock minimisation should trivially be the occupation of the $\eta$ lowest energy spin-orbitals, where $\eta=N_s/2$ (for the hydrogen atom chains) is the number of electrons. However if the molecular orbitals were not pre-optimised (for example, the use of alternate bases can reduce the number of terms in the Hamiltonian, see Appendix \ref{sec:scaling}) then the optimal trial circuit parameters will need to be found. The trial circuit used here is based on that used in \cite{google_HF}.


From the initial state, a series of parameterised Givens rotations \cite{wecker_SolvingElectronModels} is applied between neighbouring qubits following the optimal layout determined in \cite{kivlichan_LinearDepth}. The resulting circuit (see Figure \ref{fig:circuit}) has depth $\mathcal{O}(N_s)$ and gate count $\mathcal{O}(N_s^2)$. The rotations used here are defined in such a way that when the gate parameter, $\theta$, goes to $0$, the Givens rotation reduces to the identity operation and therefore a circuit with the parameter set $(0,0,0,\dots)$ will return the classically pre-optimised Hartree-Fock state.

\subsection{Problem scaling}\label{sec:scaling}
\begin{figure*}[t]
\begin{center}
\includegraphics[width=8cm]{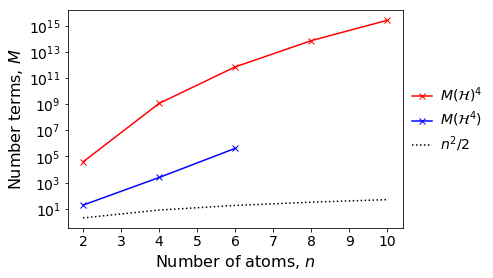}
\caption{Scaling of the number of terms in the fourth moment for the one-dimensional hydrogen chains. The blue line is the observed number of terms in $\mathcal{H}^4$ (before grouping commuting operators) while the red line represents the worst-case scenario of raising the number of terms in $\mathcal{H}$ to the power 4. The dotted black line is the number of measurements required by the method described in Appendix \ref{sec:scaling}, $n^2/2$ where $n$ is the number of atoms.
}
\label{fig:scaling}
\end{center}
\end{figure*}
While the QCM method provides a technique for the suppression of noise for VQE, which in turn was developed to take advantage of near-term quantum hardware while reducing the impact of errors, a possible limitation of VQE and, by extension, the QCM method is the potentially rapid growth of the number of terms present in the Hamiltonian which determines the number of measurements and hence the number of state preparations required of the quantum processor. In the usual molecular orbital basis the number of terms in the quantum chemical Hamiltonian scales as $\mathcal{O}(N_s^4)$ (where $N_s$ is the number of spin-orbitals) which is then amplified by the exponentiation required for the moments correction. The scaling problem in VQE has led to various methods designed to reduce the number of measurements required. These can be split into two categories. The first class of methods are those that aim to reduce the number of terms in the Hamiltonian, for example by transforming the operator into a different basis in which the number of terms scales better than the usual $\mathcal{O}(N_s^4)$ \cite{babbush_PlaneWaves,babbush_1QPlaneWaves,kottmann_BasisSetFree} or by neglecting Hamiltonian terms with a weight below a certain threshold, such as interaction terms between spatially separated wavefunctions (orbitals). The second class of methods are those that aim to reduce the number of measurements (but not the number of terms) by partitioning the Hamiltonian terms into groups of mutually commuting, and therefore simultaneously measurable, operators. There are a variety of methods through which this grouping can be performed such as mapping to a minimum clique cover problem \cite{verteletskyi_TPBMinimumCliqueCover} or using an identity-operator sorting algorithm \cite{vallury_QCM}. 


In this work, we make use of the fact that only the single-excitation operators need to be measured. For a system represented by $N_s$ spin-orbitals this already limits the number of expectation value measurements to $\mathcal{O}(N_s^2)$. A refinement performed in \cite{google_HF} further reduces the necessary number of measurements to $N_s/2+1$. Here, since the exponentiation of the Hamiltonian scales as $\mathcal{O}(N_s^{16})$ at worst, the $\mathcal{O}(N_s)$ method is not performed and an alternative that scales as $\mathcal{O}(N_s^2)$ but that avoids any need for virtual swapping or circuit recompilation, is employed instead. In comparison to the cost of calculating the moments formulae, the $\mathcal{O}(N_s^2)$ measurement cost is deemed acceptable.

To form TPB groupings the 2-mode operators are first transformed to qubit operators using the Jordan-Wigner transform \cite{JordanWigner_JWTransform}. For real-valued orbitals and $j<k$;
\begin{align}
\langle a^\dagger_ja_k\rangle &=\frac{1}{4}\big(\big\langle \dots I_{j-1}X_jZ_{j+1}\dots Z_{k-1}X_kI_{k+1}\dots\big\rangle \n&~~~~+\big\langle \dots I_{j-1}Y_jZ_{j+1}\dots Z_{k-1}Y_kI_{k+1}\dots\big\rangle \big).
\end{align}
For the $j>k$ case, $j$ and $k$ can be exchanged on the right-hand side of the equation. Each 2-mode operator can be characterised by a distance, $d=|j-k|$. By measuring the Pauli strings consisting of $X$ on every qubit and $Y$ on every qubit, all distance, $d=1$, operators can be reconstructed. Likewise measuring the strings $XZXZ\dots$, $YZYZ\dots$, $ZXZX\dots$ and $ZYZY\dots$ allows for the reconstruction of all distance, $d=2$, operators. Continuing this pattern, it can be shown that every $d\in(0,N_s/4]$ operator can be reconstructed by measuring $2d$ strings. For operators with $d\in[N_s/4,N_s/2]$ the operators are less local but there are fewer possible operators and the number of string measurements required is $N_s-2d$. For the cases considered here, $N_s/4$ is always an integer since the singlet state requires an even number of hydrogen atoms and the spin degeneracy introduces a further factor of 2. In the special case of $d=0$ the expectation value becomes $\langle a^\dagger_ja_j\rangle =\frac{1}{2}(1-\langle Z_j\rangle )$ and can be reconstructed from the four measurements made for the $d=2$ case. The total number of measurements required is then,
\begin{align}
M&=\sum_{d=1}^{N_s/4-1}2d+\sum_{d=N_s/4}^{N_s/2-1}(N_s-2d)=\frac{N_s^2}{8},\n &\hspace{8ex}(N_s>4,~N_s/4\in\mathbb{Z}).
\end{align}
Compared to the naive method of measuring each term in the moments formulae individually which scales at worst as $\mathcal{O}(N_s^{16})$, the $\mathcal{O}(N_s^2)$ cost is a significant improvement.

\subsection{RDM purification}\label{sec:purification}
In the Hartree-Fock approximation, the one-body reduced density matrix (1-RDM), $R$, is expected to be idempotent ($R=R^2$). In practice, the presence of errors in the application of gates or in readout, as well as the statistical effects of shot noise lead to the calculated 1-RDM being only nearly idempotent. In this work, the McWeeny iterative purification method \cite{mcweeny_DensityMatrices,google_HF} is used to correct for these errors. A brief motivation of the procedure is given below;

The error in the RDM (in the sense of its distance from idempotency) can be quantified by $D=R^2-R$ in which case the objective of the purification is to reduce all elements of the matrix $D$ to 0. By framing this as a minimisation problem with cost function
\begin{align}
C=\sum_{ij}(D_{ij})^2\nonumber
\end{align}
the RDM can be purified by gradient descent with a step size $\delta$ according to;
\begin{align}
R'&=R-\delta\nabla_R(C),\label{eqn:GradientDescent}
\end{align}
where $R$ and $R'$ are the current and updated 1-RDMs respectively and $\nabla_R(C)$ is the gradient of the cost function with respect to the matrix $R$;
\begin{align}
\nabla_R(C)&=\left[\matrix{
$\frac{\partial C}{\partial R_{11}}$&$\frac{\partial C}{\partial R_{12}}$&$\cdots$\\
$\frac{\partial C}{\partial R_{21}}$&$\frac{\partial C}{\partial R_{22}}$&$\cdots$\\
$\vdots$&$\vdots$&$\ddots$
}\right].\nonumber
\end{align}

In the case of real-valued orbitals, the error matrix $D$ inherits the symmetry of the 1-RDM and so the cost function can be rewritten using the identity
\begin{align}
\sum_{ij}(D_{ij})^p&={\rm Tr}(D^p),\label{eqn:MatrixElementPowerToTrace}
\end{align}
for symmetric matrix $D$ and non-negative integer $p$:
\begin{align}
C&={\rm Tr}(R^4-2R^3+R^2).\nonumber
\end{align}
So the gradient evaluates to
\begin{align}
\nabla_R(C)&=4R^3-6R^2+2R.\nonumber
\end{align}
Substituting this into \ref{eqn:GradientDescent} and choosing the step size to be $\delta=1/2$, the expected McWeeny purification formula is recovered
\begin{align}
R'=3R^2-2R^3.\label{eqn:McWeenyPurification}
\end{align}
Once the 1-RDM has been measured
\begin{align}
R&=\left[\matrix{
$\langle a^\dagger_0a_0\rangle$&$\langle a^\dagger_0a_1\rangle$&$\cdots$\\
$\langle a^\dagger_1a_0\rangle$&$\langle a^\dagger_1a_1\rangle$&$\cdots$\\
$\vdots$&$\vdots$&$\ddots$\\
}\right]
\end{align}
equation \ref{eqn:McWeenyPurification} can be applied iteratively to obtain a corrected 1-RDM that more accurately represents a single Slater-determinant state. Using the results of Section \ref{sec:reduction}, the energy can then be calculated from the elements of this corrected 1-RDM.

\subsection{Sample measurement data}\label{sec:SIdata}
Data for the H$_2$ molecular Hamiltonian calculated at a bond length of 0.74\angstrom using the minimal STO-3G basis is presented in table 1. Table 2 contains sample data from the quantum processor \emph{ibmq\_sydney} and from which the 1-RDM for molecular hydrogen (with circuit parameter $\theta=0$) can be constructed. Figure SI.2 presents estimated energies for the H$_4$ chain for 97 sets of random circuit parameters for both direct and moments-based measurement.


\begin{table*}[tbp]
\begin{center}
\begin{tabular}{|c|c|c|c|c|}\hline
Fermionic operator&$\langle\mathcal{H}\rangle$&$\langle\mathcal{H}^2\rangle$&$\langle\mathcal{H}^3\rangle$&$\langle\mathcal{H}^4\rangle$\\\hline\hline
$a^\dagger_0a_0$&-1.2533&1.5708&-1.9687&2.4674\\\hline
$a^\dagger_1a_1$&-0.4751&0.2257&-0.1072&0.0509\\\hline
$a^\dagger_2a_2$&-1.2533&1.5708&-1.9687&2.4674\\\hline
$a^\dagger_3a_3$&-0.4751&0.2257&-0.1072&0.0509\\\hline
$a^\dagger_1a^\dagger_0a_1a_0$&-0.4825&0.2443&-0.142&0.1089\\\hline
$a^\dagger_2a^\dagger_0a_2a_0$&-0.6748&-0.247&2.3385&-6.6903\\\hline
$a^\dagger_2a^\dagger_0a_3a_1$&-0.1812&0.3777&-0.7094&1.3164\\\hline
$a^\dagger_2a^\dagger_1a_2a_1$&-0.6637&0.6301&-0.7642&1.009\\\hline
$a^\dagger_2a^\dagger_1a_3a_0$&-0.1812&0.3859&-0.6222&0.9001\\\hline
$a^\dagger_3a^\dagger_0a_2a_1$&-0.1812&0.3859&-0.6222&0.9001\\\hline
$a^\dagger_3a^\dagger_0a_3a_0$&-0.6637&0.6301&-0.7642&1.009\\\hline
$a^\dagger_3a^\dagger_1a_2a_0$&-0.1812&0.3777&-0.7094&1.3164\\\hline
$a^\dagger_3a^\dagger_1a_3a_1$&-0.6977&0.3548&-0.1216&-0.0501\\\hline
$a^\dagger_3a^\dagger_2a_3a_2$&-0.4825&0.2443&-0.142&0.1089\\\hline
$a^\dagger_2a^\dagger_1a^\dagger_0a_2a_1a_0$&&1.3926&-3.913&8.7428\\\hline
$a^\dagger_3a^\dagger_1a^\dagger_0a_3a_1a_0$&&0.6637&-1.1088&1.4847\\\hline
$a^\dagger_3a^\dagger_2a^\dagger_0a_3a_2a_0$&&1.3926&-3.913&8.7428\\\hline
$a^\dagger_3a^\dagger_2a^\dagger_1a_3a_2a_1$&&0.6637&-1.1088&1.4847\\\hline
$a^\dagger_3a^\dagger_2a^\dagger_1a^\dagger_0a_3a_2a_1a_0$&&2.4195&-5.4784&10.9158\\\hline
\end{tabular}
\caption{Fermionic operator weights for molecular hydrogen calculated in the STO-3G basis at equilibrium bond length (0.74\angstrom). Molecular orbitals are ordered from 0 to 3 as spin-up bonding, spin-up anti-bonding, spin-down bonding, spin-down anti-bonding}
\end{center}
\end{table*}

\begin{table*}[tbp]
\begin{center}

\begin{tabular}{cc}
\begin{tabular}{c|c|c|}
$\langle a^\dagger_ia_j\rangle$&$j=0,2$   &$j=1,3$ \\\hline
$i=0,2$                          & 0.96   &-0.0114 \\\hline
$i=1,3$                          &-0.0114 & 0.0236 \\\hline
\end{tabular}
&
\begin{tabular}{c|c|c|}
$\langle a^\dagger_ia_j\rangle$&$j=0,2$   &$j=1,3$ \\\hline
$i=0,2$                          & 0.9619 &-0.0082 \\\hline
$i=1,3$                          &-0.0082 & 0.0229 \\\hline
\end{tabular}
\\
\\
\begin{tabular}{c|c|c|}
$\langle a^\dagger_ia_j\rangle$&$j=0,2$   &$j=1,3$ \\\hline
$i=0,2$                          & 0.9647 &-0.0036 \\\hline
$i=1,3$                          &-0.0036 & 0.023  \\\hline
\end{tabular}
&
\begin{tabular}{c|c|c|}
$\langle a^\dagger_ia_j\rangle$&$j=0,2$   &$j=1,3$ \\\hline
$i=0,2$                          & 0.9675 &-0.0095 \\\hline
$i=1,3$                          &-0.0095 & 0.0262 \\\hline
\end{tabular}
\end{tabular}
\caption{Four sets of 1-RDM elements for the hydrogen molecule ($\theta=0$) calculated on \emph{ibmq\_sydney} with $\approx2000$ shots each. The ordering of orbitals matches that used in table 1.}
\end{center}
\end{table*}

\begin{figure*}[tb]
    \centering
    \includegraphics[width=0.6\linewidth]{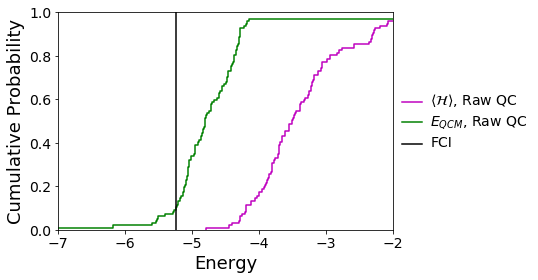}
    \caption{Cumulative frequency plot for the H$_4$ molecular chain. 100 sets of uniform random parameters were chosen and the energy was evaluated both directly and by the QCM method on the quantum processor \textit{ibmq\_montreal}. The vertical axis is the cumulative frequency, i.e. the fraction of points with a measured energy below a given value. In the ideal case (that the FCI energy is recovered regardless of the parameter values) the resulting plot would be the step function at the FCI energy (black line). Of the 100 random points, only 97 are represented in the plot. The remaining 3 points were rejected due to large variations in the QCM energy when repeated, e.g. standard deviations of the same scale or larger than the energy measurements. These anomalous results are likely due to the combination of a poorly conditioned point and errors in the device leading to non-physical density matrices. The large deviations are seen to vanish when McWeeny purification is applied.
    }
    \label{fig:frequency}
\end{figure*}

\end{multicols}

\end{document}